\documentclass[fleqn]{2023SCGE}
\usepackage{acro}

\DeclareAcronym{SGWB}{
  short = SGWB ,
  long = stochastic gravitational-wave background ,
  short-plural = s ,
}

\DeclareAcronym{CMB}{
  short = CMB ,
  long = cosmic microwave background ,
  short-plural =  ,
}

\DeclareAcronym{PTA}{
  short = PTA ,
  long = pulsar timing array ,
  short-plural = s ,
}

\DeclareAcronym{LISA}{
  short = LISA ,
  long = Laser Interferometer Space Antenna ,
  short-plural =  ,
}

\DeclareAcronym{PBH}{
  short = PBH ,
  long = primordial black hole ,
  short-plural = s ,
}

\DeclareAcronym{SIGW}{
  short = SIGW ,
  long = scalar-induced gravitational wave ,
  short-plural = s ,
}
\DeclareAcronym{IGW}{
  short = IGW ,
  long = inflationary gravitational wave ,
  short-plural = s ,
}
\DeclareAcronym{PGW}{
  short = PGW ,
  long = preheating gravitational wave ,
  short-plural = s ,
}
\DeclareAcronym{FOPT}{
  short = FOPT ,
  long = first-order phase transition ,
  short-plural = s ,
}
\DeclareAcronym{CV}{
  short = CV ,
  long = cosmic variance ,
  short-plural =  ,
}
\DeclareAcronym{SNR}{
  short = SNR ,
  long = signal-to-noise ratio ,
  short-plural = s ,
}
\begin{document}

\ensubject{subject}

\ArticleType{Article}
\SpecialTopic{SPECIAL TOPIC: }
\Year{2023}
\Month{January}
\Vol{66}
\No{1}
\DOI{??}
\ArtNo{000000}
\ReceiveDate{}
\AcceptDate{}

\title{Measuring the anisotropies in astrophysical and cosmological gravitational-wave backgrounds with Taiji and LISA networks}{Measuring the anisotropies in astrophysical and cosmological gravitational-wave backgrounds with Taiji and LISA networks}

\author[1]{Zhi-Chao Zhao}{}%
\author[2]{Sai Wang}{{wangsai@ihep.ac.cn}}

\AuthorMark{Z.-C. Zhao}

\AuthorCitation{Z.-C. Zhao, S. Wang}

\address[1]{Department of Applied Physics, College of Science, China Agricultural University, 17 Qinghua East Road, Haidian District, Beijing 100083, China}
\address[2]{Theoretical Physics Division, Institute of High Energy Physics, Chinese Academy of Sciences, Beijing 100049, China}


\abstract{We investigate the capabilities of space-based gravitational-wave detector networks, specifically Taiji and LISA, to measure the anisotropies in stochastic gravitational-wave background (SGWB), which are characterized by the angular power spectrum. We find that a detector network can improve the measurement precision of anisotropies by at most fourteen orders of magnitude, depending on the angular multipoles. By doing so, we can enhance our understanding of the physical origins of SGWB, both in astrophysical and cosmological contexts. We assess the prospects of the detector networks for measuring the parameters of angular power spectrum. We further find an inevitable effect of cosmic variance, which can be suppressed by a better angular resolution, strengthening the importance of configuring detector networks. Our findings also suggest a potential detection of the kinematic dipole due to Doppler boosting of SGWB.}



\maketitle


\begin{multicols}{2}
\section{Introduction}\label{sec:intro}

Identifying the origin of \ac{SGWB} is one of the most important research projects within the community \cite{Christensen:2018iqi,LISACosmologyWorkingGroup:2022jok}. 
Generally, there are two origins, which are either astrophysical or cosmological. 
The former stands for an incoherent superposition of unresolved individual sources, e.g., binary black holes, binary neutron stars, and so on (e.g., see reviews in Ref.~\cite{Regimbau:2011rp}). 
The latter might consist of a variety of gravitational-wave sources in the early Universe, such as the topological defects, first-order phase transition, enhanced curvature perturbations, preheating, inflationary quantum fluctuations, and so on (e.g., see reviews in Ref.~\cite{Caprini:2018mtu}).
The cosmological \ac{SGWB} plays as a 
\Authorfootnote

\noindent 
probable probe of the high-energy physics beyond the capability of ongoing and planned colliders. 
However, it is challenging to disentangle the cosmological origin from the astrophysical one, particularly when they share the same energy-density spectrum. 
Recently, upper limits on the energy-density spectral amplitude of \ac{SGWB} have already been reported at high frequency band \cite{KAGRA:2021kbb}. 
\Acp{PTA} found significant evidence for a \ac{SGWB} at very low frequency band \cite{Xu:2023wog,EPTA:2023fyk,NANOGrav:2023gor,Reardon:2023gzh}. 
At the lowest frequency band, the tensor-to-scalar ratio has been constrained via measurements of B-mode polarization in the \ac{CMB} \cite{BICEP:2021xfz}. 
Additionally, more and more observational data are anticipated \cite{Renzini:2022alw}.

Statistics of the anisotropies in \ac{SGWB} energy density encodes vital information of gravitational-wave sources, leading to possible identification of the origin of \ac{SGWB} (e.g., see reviews in Ref.~\cite{LISACosmologyWorkingGroup:2022kbp,LISACosmologyWorkingGroup:2022jok}). 
The angular power spectrum is the lowest-order statistical quantity, which corresponds to the angular two-point correlation function. 
Recently, the anisotropic \ac{SGWB} has been searched for at high frequency band \cite{KAGRA:2021mth} and very low frequency band \cite{NANOGrav:2023tcn}, respectively. 
Nevertheless, no significant evidence has been found, and only upper bounds on the amplitude of angular power spectrum have been reported. 
Even though null detection, the angular power spectrum is expected to serve as a powerful tool to identify the origin of \ac{SGWB} \cite{LISACosmologyWorkingGroup:2022kbp}. 
In fact, for different sources, the angular power spectra can share different dependence on the angular multipoles or frequencies or both. 
Therefore, in order to extract the useful statistical information, it is essential to study the angular sensitivities of gravitational-wave detectors.

In this work, we explore the detection prospects for the angular power spectrum of \ac{SGWB} via analyzing the angular sensitivities of space-based detector networks composed of Taiji \cite{Hu:2017mde} and \ac{LISA} \cite{Colpi:2024xhw}, or a dual Taiji. 
The angular sensitivity of \ac{LISA} has been studied and the angular resolution is limited to detect only a few angular multipoles with the lowest $\ell$'s \cite{LISACosmologyWorkingGroup:2022kbp}. 
Recently, the Taiji+\ac{LISA} detector network has been proposed, which can significantly improve the precision for localization of an individual source \cite{Ruan:2020smc} and measurements of anisotropies \cite{Cui:2023dlo,Mentasti:2023uyi}. 
We will study the angular sensitivity of such a network. 
As a fantasy, we will propose a new detector network composed of a dual Taiji, i.e., the Taiji+Taiji network, with \ac{LISA} being replaced with a second Taiji. 
We will study to what extent the detector networks can enhance the angular sensitivity, compared with an individual detector. 
We will obtain the noise power spectral density with respect to the angular multipoles, which are now extended to higher $\ell$'s. 
Further, we will study the detection threshold for the angular power spectrum, and the possibility of detection of kinematic dipole arising from the motion of the Solar System relative to \ac{SGWB}. 
In particular, we investigate the inevitable effect of cosmic variance. 
In order to keep our study to be generic, we will not focus on a specific source of \ac{SGWB}, but refer to a framework of parameterization, as demonstrated in the context. 
However, specific sources can be straightforwardly studied via following the same method or adjusting our results appropriately.

The remaining context of the paper is as follows. 
We briefly summarize knowledge of the angular power spectrum of \ac{SGWB} in Section~\ref{sec:apsgw}. 
We analyze the angular sensitivities of detector networks in Section~\ref{sec:asdet}. 
We study the detection prospects for the anisotropies in \ac{SGWB}, including the kinematic dipole, in Section~\ref{sec:dproa}. 
Conclusions and discussion are shown in Section~\ref{sec:condi}. 
Throughout this paper, we adopt cosmological parameters in \texttt{Planck} 2018 results as our fiducial model.

\section{Angular power spectrum of SGWB}\label{sec:apsgw}

We survey the basic knowledge of the anisotropies in \acp{SGWB} of both the astrophysical and cosmological origins. 
We briefly demonstrate both the angular multipole and frequency dependence of the angular power spectra of these \acp{SGWB}, but leave details to the literature (e.g., see Ref.~\cite{LISACosmologyWorkingGroup:2022kbp}). 
When considering the most sensitive bands of Taiji and \ac{LISA}, i.e., $\sim \mathcal{O}(1)$ milli-Hertz, we can approximate the energy-density fraction spectrum $\Omega_{\mathrm{GW}}$ and angular power spectrum $C_{\ell}$ of a given \ac{SGWB} to be power-laws. 
To be specific, the index of energy-density spectrum is constant $\alpha$, and that of angular power spectrum is constant $\gamma$. 
However, our research approach can be straightforwardly generalized to other types of frequency dependence if needed. 
Furthermore, since the angular resolutions of detectors are limited, we focus on the low-$\ell$ angular multipoles of \ac{SGWB} anisotropies.

We consider the astrophysical \ac{SGWB} produced from inspiralling binary compact objects, e.g., binary black holes, binary neutron stars, and so on (e.g., see Ref.~\cite{Regimbau:2011rp}). 
Generally, these sources trace the large-scale distribution of non-relativistic matter in the Universe. 
For the anisotropies in \ac{SGWB}, the angular power spectrum has an $\ell$ dependence, roughly scaling as $C_{\ell}\propto (\ell+1/2)^{-1}$ \cite{Cusin:2018rsq, Cusin:2019jhg,Cusin:2017fwz,  Jenkins:2018kxc, Jenkins:2018uac, Contaldi:2016koz, Wang:2021djr, Bavera:2021wmw, Jenkins:2019nks, Mukherjee:2019oma, Bellomo:2021mer,Cusin:2019jpv}. 
It also has a frequency dependence with power-law index $2/3$. 
Furthermore, it is known that the shot noise is larger by around two orders of magnitude than the anticipated signal at the high-frequency band \cite{Jenkins:2019uzp,Jenkins:2019nks,Cusin:2019jpv,Alonso:2020mva,Mukherjee:2019oma,Bellomo:2021mer,Canas-Herrera:2019npr}, leading the observations to be particularly challenging. 
In contrast, the shot noise is negligible at low-frequency band \cite{Canas-Herrera:2019npr,Cusin:2019jhg,Scelfo:2021fqe,Capurri:2021zli}. 
This is one of the most important reasons that we focus on the space-based detectors, rather than the ground-based ones.

There are a quantity of cosmological sources for \ac{SGWB} (e.g., see Ref.~\cite{Caprini:2018mtu}). 
In this work, we consider \acp{SGWB} produced from the inflationary quantum fluctuations, enhanced curvature perturbations, \acp{FOPT}, and cosmic domain walls, since their anisotropies are potentially detectable for Taiji and \ac{LISA} \cite{LISACosmologyWorkingGroup:2022kbp,LISACosmologyWorkingGroup:2022jok}. 
Other sources are disregarded in this work, but they can be studied in the same way if needed.

\Acp{IGW} originate from the quantum fluctuations during an inflation epoch of the early Universe \cite{Starobinsky:1979ty,Starobinsky:1980te,Guth:1980zm,Sato:1980yn,1982PhLB..108..389L,Albrecht:1982wi}. 
When assuming the primordial non-Gaussianity to be direction independent and scale invariant, the angular power spectrum has roughly an $\ell$ dependence of the form $C_{\ell}\propto [\ell(\ell+1)]^{-1}$ \cite{Adshead:2020bji, Dimastrogiovanni:2021mfs, Dimastrogiovanni:2019bfl, Jeong:2012df, ValbusaDallArmi:2023nqn}. 
In addition, the spectral amplitude strongly depends on the amplitude of primordial non-Gaussianity, indicating that the anisotropies in \ac{IGW} background are a probe of the inflationary dynamics.

\Acp{SIGW} arise from nonlinear couplings of the enhanced linear curvature perturbations \cite{Baumann:2007zm,Espinosa:2018eve,Kohri:2018awv,Ananda:2006af,Adshead:2021hnm,Ragavendra:2021qdu,Li:2023qua,Yuan:2023ofl,Li:2023xtl,Perna:2024ehx}, which are believed to gravitationally collapse into \acp{PBH} (e.g., see reviews in Ref.~\cite{LISACosmologyWorkingGroup:2023njw}). 
Large anisotropies can be produced by couplings between modes of short and long wavelengths due to the primordial non-Gaussianity of local type. 
The angular power spectrum is shown to have an $\ell$ dependence, which roughly scales as $C_{\ell}\propto [\ell(\ell+1)]^{-1}$ \cite{Bartolo:2019zvb,Li:2023qua, Wang:2023ost,Li:2023xtl,Yu:2023jrs, Dimastrogiovanni:2022eir}. 
The spectral amplitude is greatly influenced by the non-Gaussian amplitude parameter, implying that the anisotropies in \ac{SIGW} background are also a probe of inflationary dynamics.

During the \acp{FOPT} in the early Universe, gravitational waves can be produced from bubble wall collisions, magnetohydrodynamic turbulence, and sound waves \cite{Steinhardt:1981ct,Kosowsky:1991ua,Cai:2017tmh}. 
If we focus on the bubble wall collisions, this \ac{SGWB} can get sizable anisotropies when the initial conditions for perturbations are isocurvature and the universe undergone a period of early matter domination. 
Therefore, the anisotropies can be used for discriminating the adiabatic and isocurvature initial conditions. 
It is shown that the angular power spectrum has an $\ell$ dependence of the form $C_{\ell} \propto [\ell(\ell+1)]^{-1}$ \cite{Li:2022svl,Li:2021iva, Domcke:2020xmn, Jinno:2021ury, Geller:2018mwu, Racco:2022bwj,Bodas:2022urf}.

Annihilation of cosmic domain walls can also give rise to the production of \ac{SGWB} (e.g., see reviews in Ref.~\cite{Saikawa:2017hiv}). 
Large-scale anisotropies of this \ac{SGWB} can be induced by the quantum perturbations of a light scalar field, rather than the inflaton, during inflation. 
They also display as a potential probe of the inflationary physics. 
It has been shown that the angular power spectrum has an $\ell$ dependence, i.e., $C_{\ell}\propto [\ell(\ell+1)]^{-1}$ \cite{Liu:2020mru}. 

Other cosmological sources can also give arise to the anisotropies in \ac{SGWB}, e.g., the parametric resonance during preheating \cite{Bethke:2013aba,Bethke:2013vca} and the cosmic strings \cite{Olmez:2011cg,Kuroyanagi:2016ugi,Jenkins:2018nty}. 
However, the peak frequency band for the former is so high that it may be beyond the sensitivity regime of Taiji and \ac{LISA}. 
The amplitude of angular power spectrum for the latter is anticipated to be so small that it is challenging to be detected by Taiji and \ac{LISA}. 
Therefore, they are disregarded in our present work.

In conclusion, for the sake of generality, we parameterize the angular power spectrum as \footnote{{\color{black} Throughout this paper, we adopt the concept of angular power spectrum that is not rescaled with the energy-density fraction spectrum $\Omega_{\mathrm{GW}}$. }}
\begin{equation}\label{Clmodel}
    C_{\ell} \simeq A_{p}\left(\ell+\frac{1}{2}\right)^{\gamma} \ ,
\end{equation}
where $A_p$ denotes the spectral amplitude at a pivot frequency $f_{p}=1$\,milli-Hertz, and we have $\gamma=-1$ for the astrophysical \acp{SGWB} while approximately $\gamma=-2$ for cosmological ones. 
We wonder if the Taiji and \ac{LISA} networks have potentials to discriminate the astrophysical and cosmological origins. 
In addition, it is known that the \ac{CV} would lead to an inevitable uncertainty $\sigma_{C_{\ell}}$ at $1\sigma$ confidence level, i.e., 
\begin{equation}\label{eq:cv}
   \frac{\sigma_{C_{\ell}}}{C_\ell}=\sqrt{\frac{2}{2\ell+1}} \ ,
\end{equation}
which will be taken into account in the following analysis. 
{\color{black} Within the frequency bands of space-based gravitational-wave detectors, the number of astrophysical gravitational-wave sources in each frequency bin will introduce uncertainties in \( C_\ell \), known as the shot noise \cite{Canas-Herrera:2019npr, Cusin:2019jhg, Scelfo:2021fqe}. However, in the frequency range we are considering, i.e., $\sim1\,$mHz, the number of astrophysical gravitational-wave sources is sufficiently large. Therefore, the shot noise is negligible \cite{Wang:2021djr,Cusin:2019jhg}. }

\section{Noise angular power spectra of detector networks}\label{sec:asdet}

\begin{figure*}[ht]
    \centering
    \includegraphics[width=\textwidth]{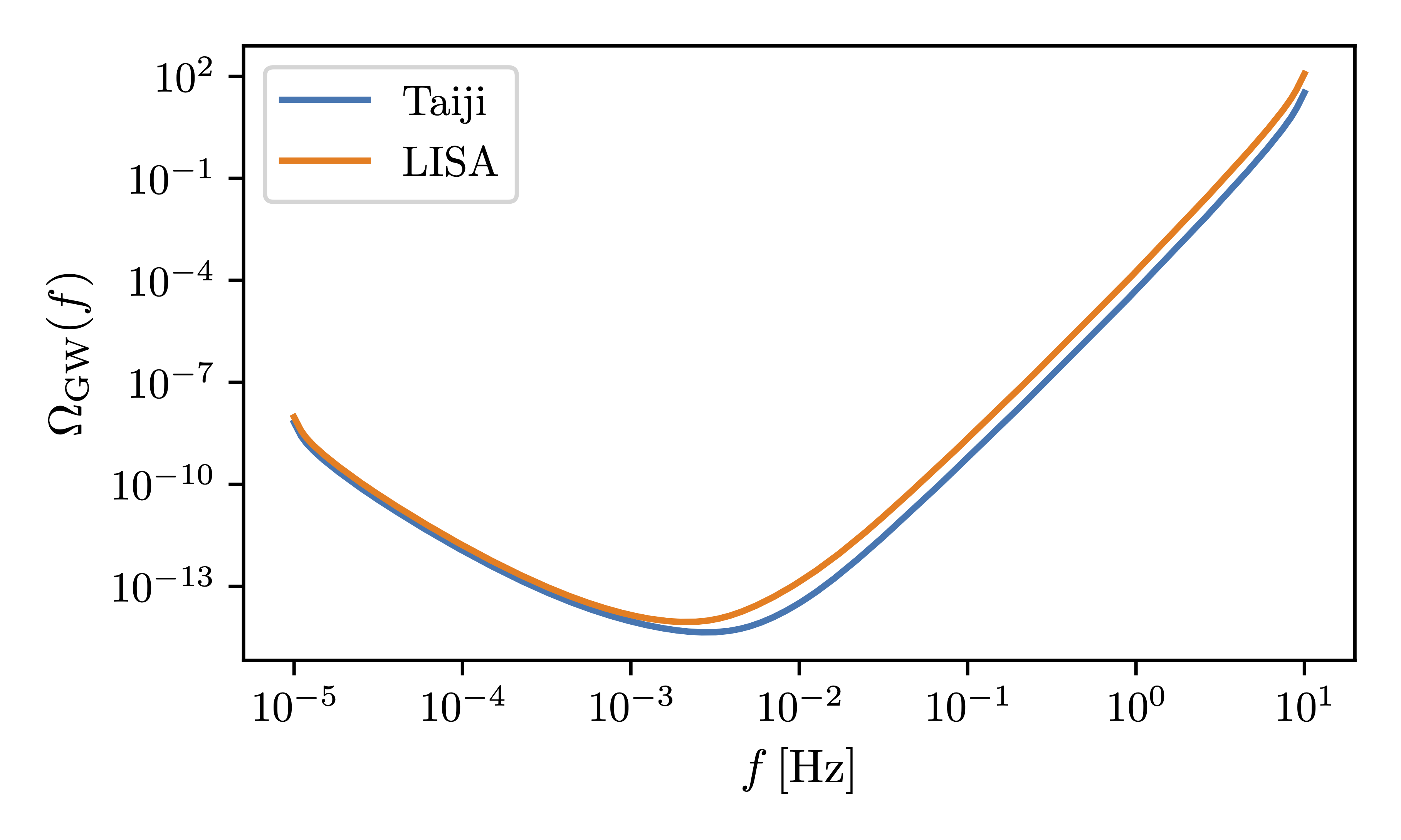}
    \caption{Power-law-integrated sensitivity curves of Taiji (blue) and LISA (orange). }
    \label{fig:sensitivities}
\end{figure*}

\begin{figure*}[ht]
    \centering
    \includegraphics[width=\textwidth]{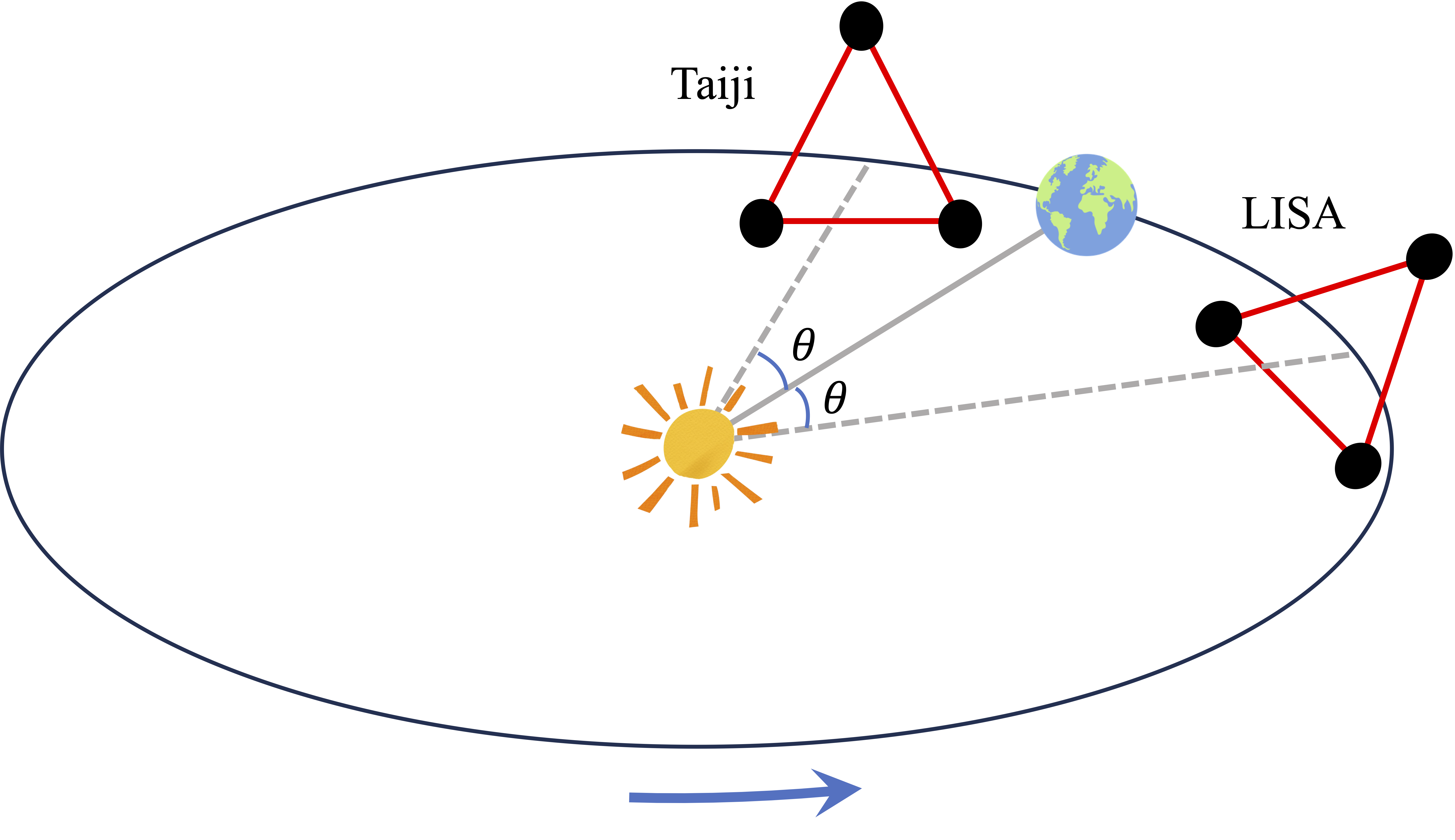}
    \caption{Spatial configuration of space-based detector networks. The separation angle between the line from the Sun to an individual detector and the line from the Sun to the Earth is $\theta=20^{\circ}$. }
    \label{fig:configuration}
\end{figure*}

Space-based gravitational-wave detectors, namely Taiji \cite{Hu:2017mde} and \ac{LISA} \cite{Colpi:2024xhw}, are planned to launch in 2030s. 
Both of them consist of three identical satellites. 
Any two satellites are separated by $3\times10^6$\,km for Taiji while $2.5 \times 10^6$\,km for \ac{LISA}. 
Such long arms let the two detectors to deserve high sensitivities to a \ac{SGWB} in the milli-Hertz band. 
In Fig.~\ref{fig:sensitivities}, we illustrate the {\color{black} 4-years} power-law-integrated sensitivity curves, which are plotted by using the publicly available \texttt{schNell} \cite{Alonso:2020rar}. Here, we fix signal-to-noise ratio (SNR) to equal $1$ for illustration. 
In addition, Taiji and \ac{LISA} share the same orbit around the Sun, as illustrated in Fig.~\ref{fig:configuration}. 
One of them is leading the Earth while the other one is behind the Earth. 
The separation angle between the line from the Sun to an individual detector and the line from the Sun to the Earth is $\theta=20^{\circ}$ \cite{Ruan:2020smc}. 
Therefore, a detector network with a baseline $\sim10^{8}$\,km can be formed, thereby leading to a possible enhancement of the angular sensitivity.

\begin{figure*}[ht]
    \includegraphics[width=\textwidth]{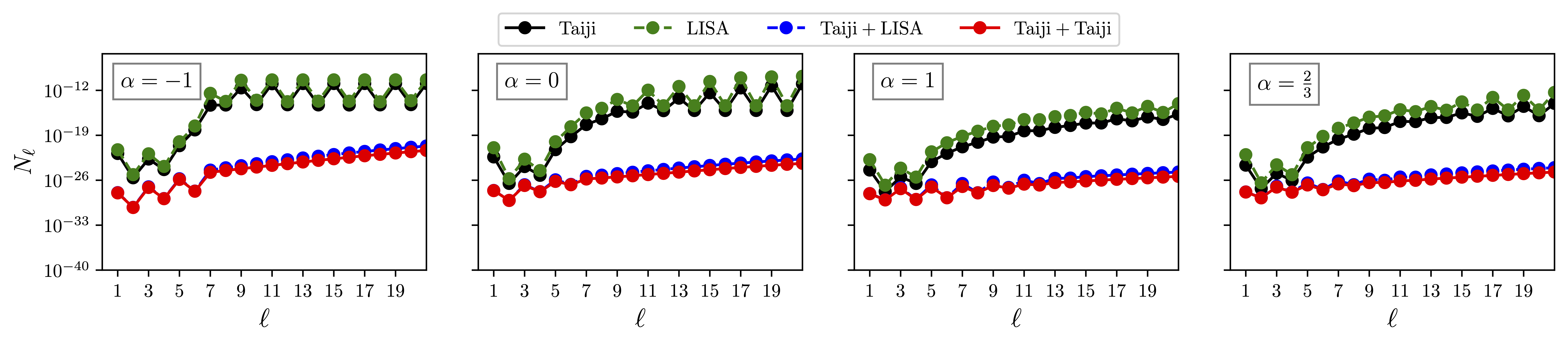}
    \caption{Noise angular power spectra of Taiji, LISA, and their networks with spatial configuration in Fig.~\ref{fig:configuration}. We consider an observing duration $T_{\mathrm{obs}}=4$\,years and four different power-laws $\alpha$ for the energy-density spectral index of SGWB. }
    \label{fig:nell}
\end{figure*}

We adopt a modified version of \texttt{schNell} \cite{Alonso:2020rar}, which has been developed to simulate the noise angular power spectrum of detectors. 
The original version already included the instrumental setup of \ac{LISA}. 
Our modified version further contains that of Taiji as well as the information of any detector network. 
In this work, we focus on the Taiji+\ac{LISA} and Taiji+Taiji networks. 
Both of them have the same spatial configuration, but for the latter, \ac{LISA} is replaced by an additional Taiji. 
Assuming the detector noise as a Gaussian random variable, we can get the noise angular power spectrum $N_\ell$, which also stands for the angular sensitivity curve. {\color{black} For readers who are interested in the detailed formulas for calculating $N_\ell$, we suggest them to refer to Ref.~\cite{Alonso:2020rar} and references therein.}

Our results of $N_\ell$ are shown in Fig.~\ref{fig:nell}. 
Here, we have considered an observing duration $T_{\mathrm{obs}}=4$\,years and four different power-laws $\alpha$ for the energy-density spectral index of \ac{SGWB}. 
{\color{black} For an individual detector, the noise $N_{\ell}$ increases sharply within $\ell\sim5-10$, due to limitation of the detector angular resolution that is determined by the length of baseline. 
In contrast, it stays almost constant for the detector networks, since the angular resolution of has been significantly enhanced due to a much longer baseline. }
On the other hand, compared with an individual detector, the detector networks dramatically suppress the noise {\color{black}for all the considered angular multipoles $\ell$}, due to better angular resolutions. 
For example, for a flat spectrum, as shown in the second panel from left to right, they suppress the noise by at most $\sim14$ orders of magnitude, indicating significant enhancements of the angular sensitivity. 
Therefore, our study has demonstrated great potential of detector networks to measure the anisotropies in \ac{SGWB}. 
Additionally, we find that the Taiji+Taiji network has a bit better sensitivity than the Taiji+\ac{LISA} network due to the longer arms of Taiji. 

\section{Detection prospects for the anisotropies in SGWB}\label{sec:dproa}

\begin{figure*}[ht]
    \includegraphics[width=\textwidth]{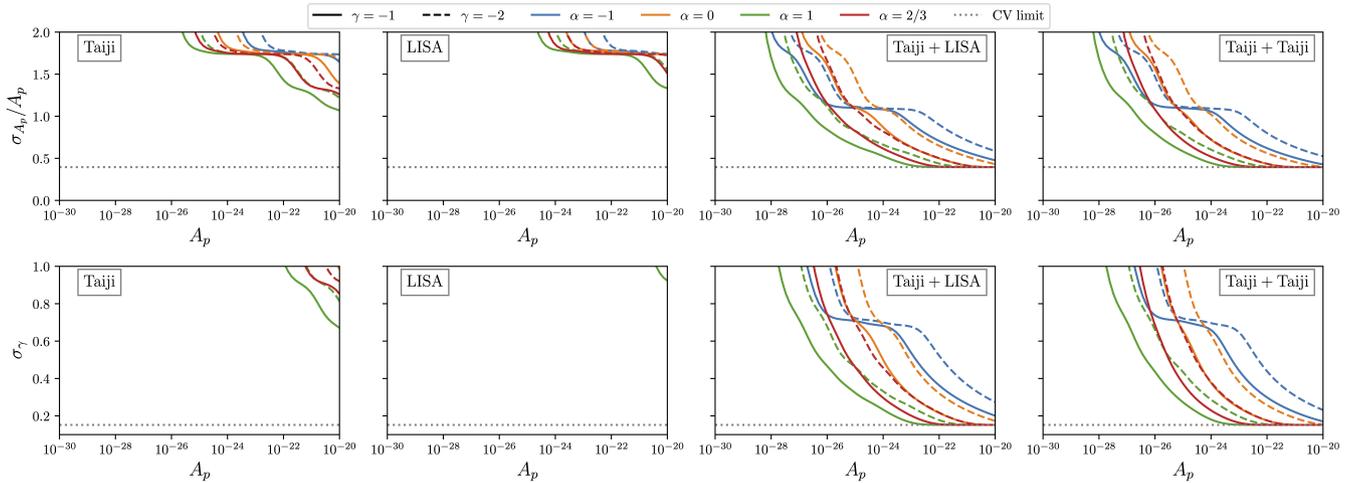}
    \caption{The $1\sigma$ uncertainties $\sigma_{A_{p}}/A_{p}$ and $\sigma_{\gamma}$ with respect to a series of fiducial $A_{p}$. Here, the fiducial value of $\gamma$ is $-2$ (solid) and $-1$ (dashed). For comparison, we display the CV limits (dotted), as shown in Fig.~\ref{fig:CVsensitivities}. }
    \label{fig:nell1}
\end{figure*}

\begin{figure*}[ht]
\centering
    \includegraphics[width=\textwidth]{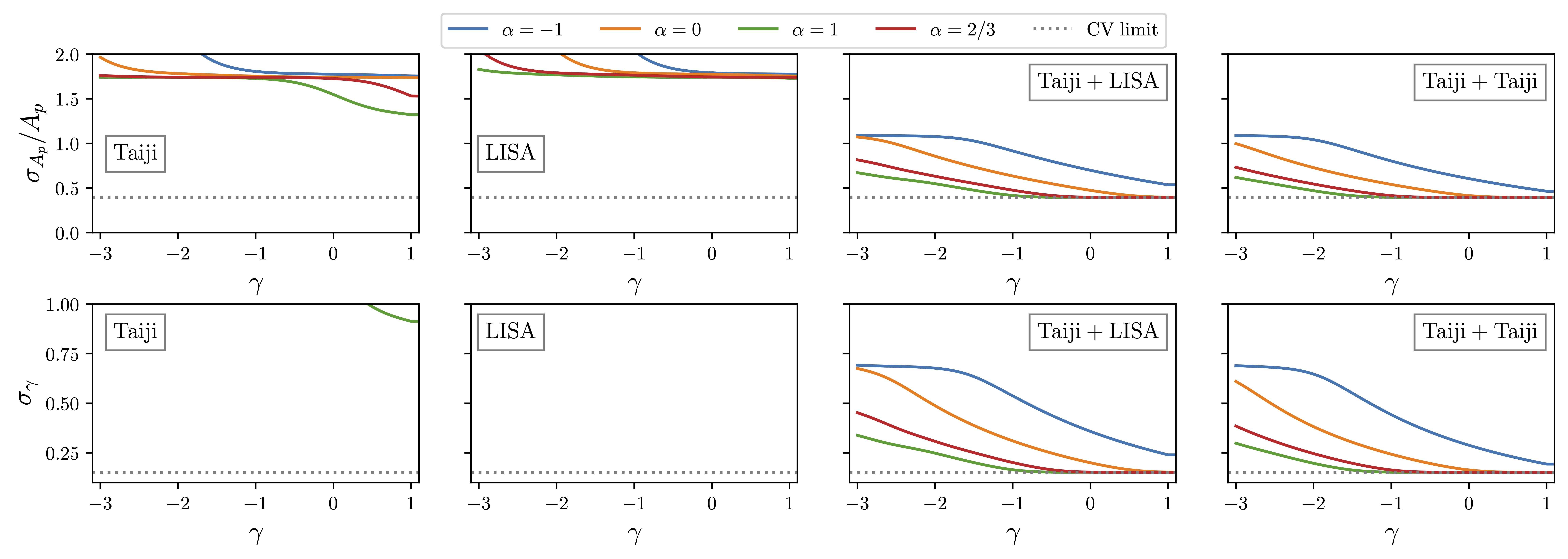}
    \caption{The $1\sigma$ uncertainties $\sigma_{A_{p}}/A_{p}$ and $\sigma_{\gamma}$ with respect to a series of fiducial $\gamma$. Here, the fiducial value of $A_{p}$ is $10^{-23}$. For comparison, we display the CV limits (dotted), as shown in Fig.~\ref{fig:CVsensitivities}. }
    \label{fig:nell2}
\end{figure*}

\begin{figure*}[ht]
\centering
    \includegraphics[width=\textwidth]{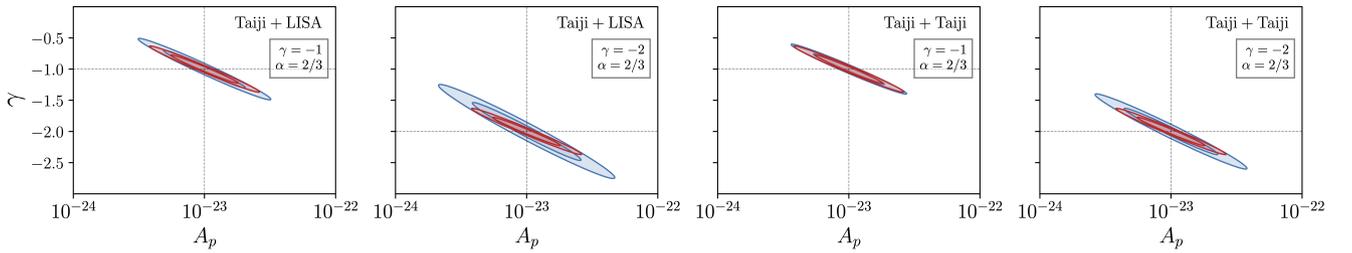}
    \caption{ The $1\sigma$ and $2\sigma$ contours for the correlation between $A_{p}$ and $\gamma$. Dotted lines denote the fiducial value. For comparison, we also show the CV limits (red regions), as shown in Fig.~\ref{fig:CVsensitivities}. }
    \label{fig:CocContour}
\end{figure*}

\begin{figure*}[ht]
    \centering
    \includegraphics[width=\textwidth]{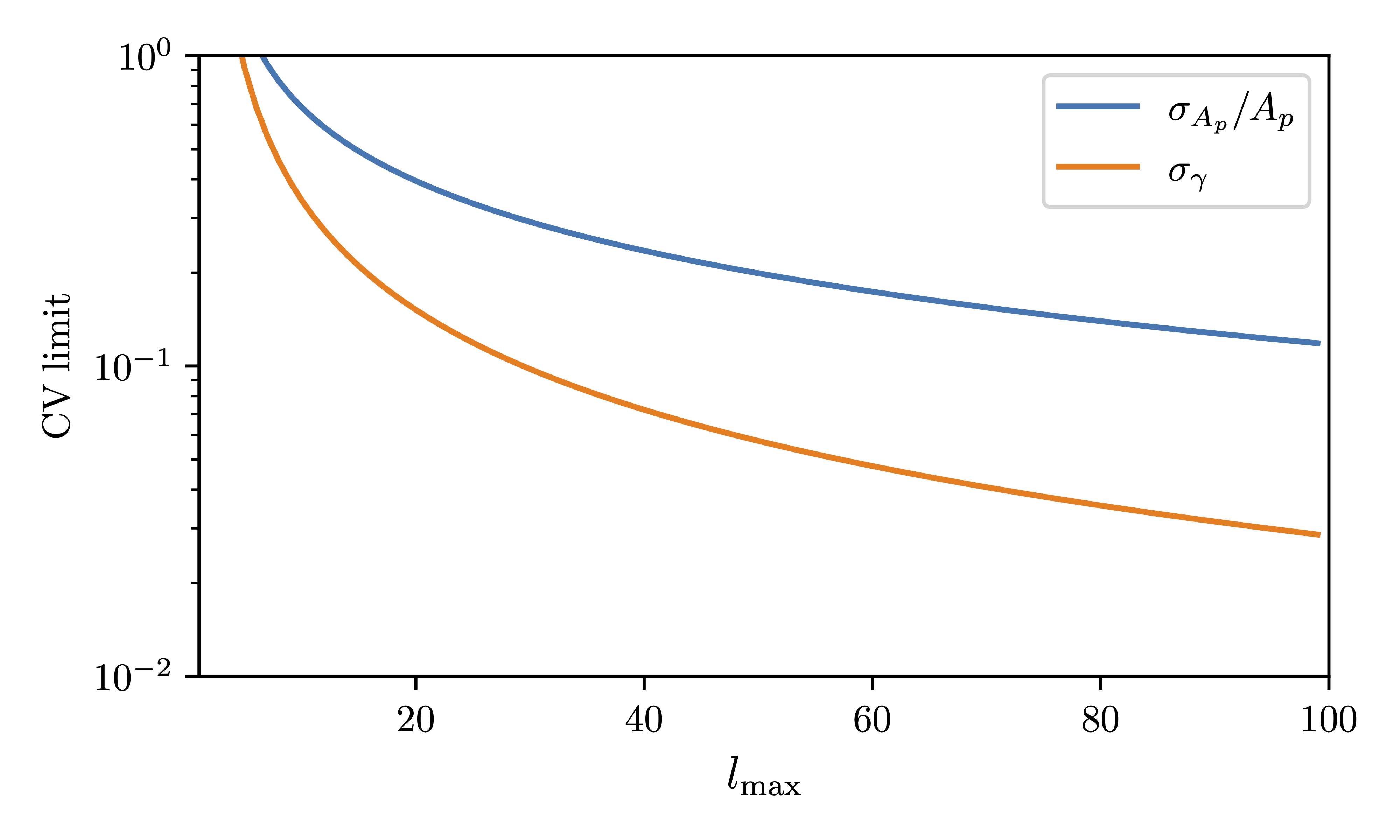}
    \caption{The CV limits on the measurements of model parameters $\ln A_{p}$ (blue) and $\gamma$ (orange). To demonstrate dependence of the CV limit on the angular resolution, we consider a series of $\ell_{\mathrm{max}}$ rather than a fixed value. }
    \label{fig:CVsensitivities}
\end{figure*}

\begin{figure*}[ht]
    \includegraphics[width=\textwidth]{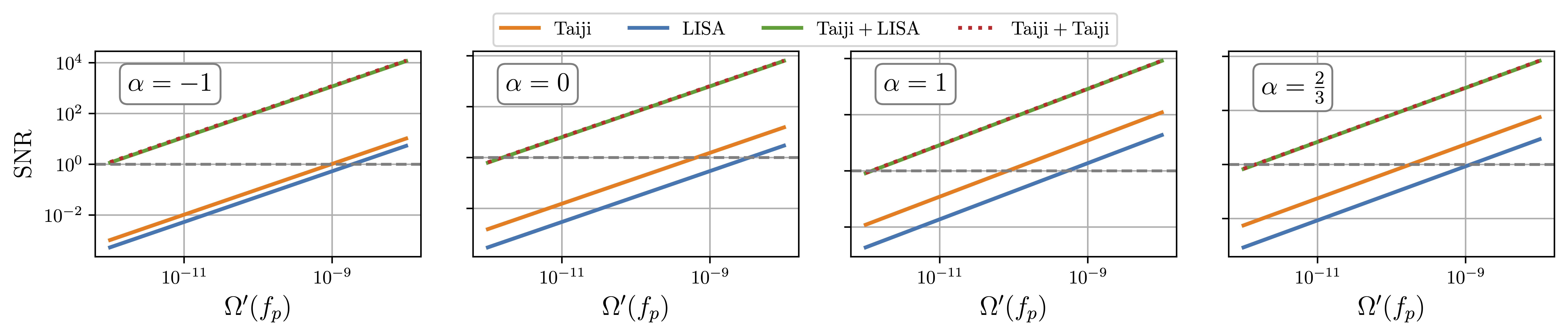}

    \caption{The same as Fig.~\ref{fig:nell}. Signal-to-noise ratio for measurements of kinematic dipole with respect to the energy-density fraction spectrum of SGWB at $f_p=1$\,milli-Hertz. {\color{black}Horizontal} dashed lines stand for SNR being unity. }
    \label{fig:snrogw}
\end{figure*}

In this section, we study the prospective sensitivities of the detector networks for measuring the anisotropies in \ac{SGWB}. 
We will forecast the uncertainties in measurements of model parameters and the inevitable effect of \ac{CV} on them. 
We will further explore potential measurements of the kinematic dipole.

\subsection{Fisher-matrix forecasting}

To conduct the forecasting, we employ the Fisher information matrix of the form 
\begin{equation}\label{eq:FisherMatrix}
F_{i j}=\sum_{\ell=2}^{\ell_{\mathrm{max}}} \frac{1}{N_\ell^2+\sigma_{C_\ell}^2} \frac{\partial C_\ell}{\partial \lambda_i} \frac{\partial C_\ell}{\partial \lambda_j}\ ,
\end{equation}
where $\lambda$ stands for a set of model parameters, i.e., $\ln{A_{p}}$ and $\gamma$, and we fix $\ell_{\mathrm{max}}=20$ for illustration unless otherwise specified. 
Here, $C_{\ell}$ and $\sigma_{C_{\ell}}$ are demonstrated in Eqs.~(\ref{Clmodel}) and (\ref{eq:cv}), respectively, and $N_{\ell}$ has been obtained via simulations in Section~\ref{sec:asdet}. 
In addition, $\ell_{\mathrm{max}}$ can be determined by the angular resolution of a given detector (network). 
Similar to the \ac{CMB}, we could have $\ell_{\mathrm{max}}\simeq180^{\circ}/\theta_{\mathrm{res}}$, where $\theta_{\mathrm{res}}$ denotes the angular resolution of detectors. 
For the Taiji+\ac{LISA} network, we might have $\theta_{\mathrm{res}}\simeq2^{\circ}$ \cite{Ruan:2020smc}, thereby leading to $\ell_{\mathrm{max}}\simeq90$. 
Our present study can be easily generalized to consider such a case, but a relevant computation would be more time-consuming.

The $1\sigma$ uncertainty of the $i$-th parameter is given by the Cramer-Rao bound, i.e., 
\begin{equation}
    \sigma_{i} = \sqrt{ \left( F^{-1} \right)_{ii} } \ ,
\end{equation}
and a correlation between the $i$-th and $j$-th parameters is $c_{ij}=(F^{-1})_{ij}/(\sigma_{i}\sigma_{j})$. 
A positive sign of $c_{ij}$ corresponds to a positive correlation, and vice versa.

Our forecasting results are summarized in Figs.~\ref{fig:nell1} and \ref{fig:nell2}. 
To be specific, we show the $1\sigma$ uncertainties $\sigma_{A_{p}}/A_{p}$ and $\sigma_{\gamma}$ with respect to a series of fiducial $A_{p}$ (the fiducial value of $\gamma$ is $-2$ and $-1$) in the former while with respect to a series of fiducial $\gamma$ (the fiducial value of $A_{p}$ is $10^{-23}$) in the latter. 
The dotted lines stand for the $1\sigma$ uncertainties contributed by \ac{CV}, as studied in the following subsection. 
For an individual detector, either Taiji or \ac{LISA}, it is impossible to measure the model parameters due to badness of its angular sensitivity. 
In contrast, a detector network can significantly change this result, leading to a possible measurement of the model parameters, depending on their fiducial value. 
If a fiducial value of $A_p$ is large enough, the measurement would be only limited by \ac{CV}, rather than the systematics. 
We should note that the above results depend on $\ell_{\mathrm{max}}$.

We further show the $1\sigma$ and $2\sigma$ contours for the correlation between $A_{p}$ and $\gamma$ in Fig.~\ref{fig:CocContour}. 
We find that $A_{p}$ and $\gamma$ are negatively correlated. 
For comparison, we also show the corresponding contours due to the cosmic variance, as studied in the following subsection. 
We further find that the angular sensitivities of the detector networks could be already comparable to the \ac{CV} limits.

\subsection{\color{black}CV limits}

If there is an ideal detector (network), i.e., $N_{\ell}=0$, we have a \ac{CV} limit on measurements of the model parameters. 
Using Eqs.~(\ref{Clmodel}), (\ref{eq:cv}), and (\ref{eq:FisherMatrix}), we obtain an analytic expression of Fisher information matrix, i.e., 
\begin{equation}
    F = \sum_{\ell=2}^{\ell_{\text{max}}} \frac{2\ell+1}{2} \begin{pmatrix}
1 & \ln\left(\ell+\frac{1}{2}\right) \\
\ln\left(\ell+\frac{1}{2}\right) & \left[\ln\left(\ell+\frac{1}{2}\right)\right]^2
\end{pmatrix}\ ,
\end{equation}
which is independent of the fiducial value of $A_p$ and $\gamma$. 
Following the same method adopted in the above subsection, we can get the inevitable uncertainties of model parameters arising from \ac{CV}.

We show the \ac{CV} limits with respect to a series of $\ell_{\mathrm{max}}$ in Fig.~\ref{fig:CVsensitivities}, where the blue and orange curves denote the \ac{CV} limits on $\sigma_{A_{p}}/{A_{p}}$ and $\sigma_{\gamma}$, respectively. 
We find that these limits decrease with an increase of $\ell_{\mathrm{max}}$, indicating that a detector (network) with better angular resolution and thereby better angular sensitivity can suppress the effect of \ac{CV} more significantly. 
Our result explicitly demonstrates the necessity of configuring the detector networks. 
To discriminate the astrophysical and cosmological origins of \ac{SGWB} at $5\sigma$ confidence level, we should consider $\ell_{\mathrm{max}}$ to be larger than $\sim15$. 
Otherwise, the \ac{CV} limit prevents such a measurement of $\gamma$. 
In contrast, we have to consider $\ell_{\mathrm{max}}\gtrsim40$ if we want to get a measurement of $A_p$ at the same confidence level, further strengthening the importance of detector networks.

\subsection{Kinematic dipole}

The motion of the solar system inevitably introduces a kinematic dipole due to the Doppler boosting of \ac{SGWB}. 
Except the monopole component, which stands for the isotropic background, the energy-density fraction spectrum of the boosted \ac{SGWB} would depend on the line-of-sight direction $\hat{\mathbf{n}}$, namely, \cite{LISACosmologyWorkingGroup:2022kbp} 
\begin{eqnarray}\label{eq:kd}
&\delta\Omega(f, \hat{\mathbf{n}})=\Omega^{\prime}(f)\bigg\{\left(\hat{\mathbf{n}} \cdot \hat{\mathbf{v}}\right) D(f) +\left[\left(\hat{\mathbf{n}} \cdot \hat{\mathbf{v}}\right)^2-\frac{1}{3}\right] Q(f)\bigg\}\ ,\nonumber\\
& 
\end{eqnarray}
where $\hat{\mathbf{v}}$ is the direction of motion of the solar system, $\Omega^{\prime}(f)$ is the energy-density fraction spectrum of the unboosted \ac{SGWB},
and we introduce 
\begin{eqnarray}
D(f) & =&\beta\left(4-n_{\Omega}\right) \ ,\nonumber\\
Q(f) & =&\beta^2\left(10-\frac{9 n_{\Omega}}{2}+\frac{n_{\Omega}^2}{2}+\frac{\alpha_{\Omega}}{2}\right)\ ,\nonumber
\end{eqnarray}
with the Doppler boosting being $\beta=0.00123$ \cite{Planck:2013kqc}, the energy-density spectral tilt and its running being defined as $n_{\Omega}(f)={d \ln \Omega^{\prime}}/{d \ln f}$ and $\alpha_{\Omega}(f)={d n_{\Omega}}/{d \ln f}$, respectively. 
In Eq.~(\ref{eq:kd}), we also consider the quadrupole, besides the dipole. 
This is because the detector (network) is more sensitive to the quadrupole than to the dipole by a few orders of magnitude, though the quadrupole is smaller than the dipole by a factor of $\beta\sim10^{-3}$. 
In the following, we still consider a constant power-law for $\Omega'(f)$, indicating $n_{\Omega}=\alpha$ and $\alpha_{\Omega}=0$.

To search for the kinematic anisotropies, a \ac{SNR} is defined as \cite{Alonso:2020rar}
\begin{equation}
    \mathrm{SNR}^2=\sum_{\ell=1}^{2}(2 \ell+1) \frac{C_{\ell}}{N_{\ell}}\ ,
\end{equation}
where we have $C_{\ell}\sim \langle\delta\Omega_{\mathrm{GW}}(f, \hat{\mathbf{n}})\delta\Omega_{\mathrm{GW}}(f, \hat{\mathbf{n}}')\rangle$ with the brackets standing for an ensemble average. 
When it is larger than unity, we would claim a measurement. 
Our results can be straightforwardly rescaled when considering other value of \ac{SNR}. 
In Fig.~\ref{fig:snrogw}, we show the results of \ac{SNR} as a function of $\Omega'(f_{p})$ for Taiji, \ac{LISA}, and their networks. 
Here, the dashed lines stand for a \ac{SNR} being unity. 
We find that compared with an individual detector, the detector networks can enhance \ac{SNR} by two or three orders of magnitude.

\section{Conclusions and discussion}\label{sec:condi}

In this work, we have investigated the potential detection prospects of Taiji and \ac{LISA} networks for future measurements of the anisotropies in both astrophysical and cosmological \acp{SGWB}. 
Our findings suggested significant improvements of both the angular resolution and angular sensitivity due to configuring the detector networks. 
In particular, we obtained the angular sensitivity curve up to tens of angular multipoles, to which an individual detector is insensitive. 
We further found that the corresponding precision of detector networks could be enhanced by at most $\sim14$ orders of magnitude, depending on the angular multipoles. 

Through analyzing the Fisher information matrix, we have assessed the measurement uncertainties of \ac{SGWB} anisotropies due to both the instrumental systematics and cosmic variance. 
In particular, we proposed the analytic formulas for calculating the effect of cosmic variance, which can not be overcome by any detector, and thereby found that it strongly depends on the measurable angular multipoles with the maximum $\ell$, indicating the importance of detector networks that have been found to significantly improve the angular resolution and angular sensitivity. 
We further found that the detector networks have potentials to discriminate the astrophysical and cosmological origins of \ac{SGWB}, as well as to measure the kinematic dipole arising from the Doppler boosting of \ac{SGWB}. 
We also showed the thresholds of $A_p$ and $\Omega'(f_{p})$ for such measurements. 

{\color{black} In this work, we assume that the Taiji and LISA detectors have an overlap observation time of 4 years. However, it is challenging to guarantee this in practice. Our calculation of \(N_\ell\) follows the method in Ref.~\cite{Alonso:2020rar}. For cases where the overlap observation time \(T\) is not 4 years, we can obtain the results through a simple transformation. In fact, \(N_\ell\) is inversely proportional to \(T\), i.e., \(N_\ell \propto 1/T\). 
To ensure that there are two detectors with the maximum overlap in observation time, we propose the Taiji+Taiji network, where if two Taiji are launched simultaneously, we can ensure at least four years of joint observation time. Furthermore, it is worthwihle to note that the sensitivity of the Taiji+Taiji network is higher than that of the Taiji+LISA network.}

Before finalizing this work, we have at least four concerns to demonstrate. 
Firstly, we have approximated both the energy-density fraction spectrum and angular power spectrum to be constant power-laws, since we focused on a narrow frequency band, which corresponds to the most sensitive band of Taiji and \ac{LISA}. 
However, our research approach can be straightforwardly generalized to study any spectral shapes and a broader frequency band, if needed. 
Secondly, though our present study has been focused on the Taiji and \ac{LISA} networks, it could be extended to detector networks composed of other detectors \cite{Jin:2023sfc,Wang:2021srv,Liang:2023fdf,Cui:2023dlo,Capurri:2022lze,Gong:2021gvw} and of other configurations \cite{Wang:2021njt,Wang:2021uih}. 
{\color{black} In particular, TianQin \cite{TianQin:2015yph} is another planned space-based detector, with an orbit around the Earth. It is expected to have an overlap observation time with LISA and Taiji. This work did not include TianQin due to the different satellite orbit for TianQin and the difficulty in guaranteeing how long the overlap observation time among the three detectors will be. However, it is expected that if the overlap observation time permits, the three-detector network, namely Taiji+TianQin+LISA, would have better detection capabilities for the anisotropies in \ac{SGWB}.}
Thirdly, {\color{black} The distribution of the binaries of white dwarfs in our own galaxy would also induce anisotropies \cite{Breivik:2019oar}, which are expected to be larger than the kinematic dipole from Doppler boosting. There have been proposals to subtract them in the literature \cite{Pieroni:2020rob,Adams:2013qma}. }
Fourthly, though only the angular power spectrum has been considered in this work, our research can be generalized to study the non-Gaussianity of \ac{SGWB} \cite{Bartolo:2019oiq,Bartolo:2019yeu,LISACosmologyWorkingGroup:2022jok,Li:2024zwx,Kumar:2021ffi,Regimbau:2011bm}, which also encodes the important statistical information of \ac{SGWB} and is characterized by the angular bispectrum and so on. 
We leave such studies to future works, since they are beyond the scope of our present paper.

\Acknowledgements{We appreciate Ming-Hui Du, Ji-Bo He, Hong-Bo Jin, Gang Wang, and Tian-Yu Zhao for helpful discussion during the final stage of preparation of this work. 
Z.C.Z. is supported by the National Key Research and Development Program of China Grant No. 2021YFC2203001. 
S.W. is supported by the National Natural Science Foundation of China (Grant No. 12175243), the National Key R\&D Program of China No. 2023YFC2206403, and the Science Research Grants from the China Manned Space Project with No. CMS-CSST-2021-B01. 
This work is supported by High-performance Computing Platform of China Agricultural University.}

\InterestConflict{The authors declare that they have no conflict of interest.}


\bibliographystyle{scpma}

\bibliography{biblio} 

\end{multicols}
\end{document}